\documentclass[prd, twocolumn, nofootinbib, floatfix]{revtex4-1}

\usepackage{hyperref}
\usepackage{amsmath}
\usepackage{graphicx}
\usepackage{dcolumn}
\usepackage{bm}
\usepackage{epsfig}
\usepackage{amssymb,latexsym,mathrsfs}
\usepackage{graphicx}
\usepackage{color}

\hypersetup{
    colorlinks=true,
    linkcolor=red,
    citecolor=blue,
    urlcolor=magenta,
} 

\newcommand{\be}{\begin{equation}}
\newcommand{\ee}{\end{equation}}
\newcommand{\bea}{\begin{eqnarray}}
\newcommand{\eea}{\end{eqnarray}}

\newcommand{\sig}{\sigma}

\begin{document}

\title{Pole Dark Energy} 

\author{Eric V.\ Linder${}^{1,2}$} 
\affiliation{
${}^1$Berkeley Center for Cosmological Physics \& Berkeley Lab, 
University of California, Berkeley, CA 94720, USA\\ 
${}^2$Energetic Cosmos Laboratory, Nazarbayev University, 
Nur-Sultan, Kazakhstan 010000 
}

\date{\today}

\begin{abstract}
Theories with a pole in the kinetic term have been used to great effect 
in studying inflation, owing to their quantum stability and attractor 
properties. We explore the use of such pole kinetic terms in dark energy 
theories, finding an interesting link between thawing and freezing models, 
and the possibility  of enhanced plateaus with ``superattractor''-like 
behavior. We assess the observational viability of pole dark energy, 
showing that simple models can give dark energy equation of state 
evolution with $w(z)<-0.9$ even for potentials that could not normally 
achieve this easily. The kinetic term pole also offers an interesting 
perspective with respect to the swampland criteria for such observationally 
viable dark energy models. 
\end{abstract} 

\maketitle

\section{Introduction}

The inflationary period of cosmic acceleration in the very early 
universe offers a rich variety of observable properties to measure, 
such as the primordial curvature perturbation power spectrum and its 
tilt $n_s$ and the primordial gravitational wave power spectrum and 
its amplitude in terms of the tensor to scalar ratio $r$. One of the exciting 
theoretical developments of the last decade is systematization of 
the relations between the two quantities, and to the number of 
inflation e-foldings $N$. Certain models, such as $\alpha$-attractors 
\cite{alpha1,alpha2,alpha3,alpha4}, give tracks, or discrete segments, 
within the $n_s$--$r$ space (see \cite{kl} for a recent review). The 
attractor nature can be traced to the pole structure of the kinetic term,  
with in turn the parameter along the track related in some instances 
to the geometry of the field space \cite{alpha3}. 

Dark energy theories also exist with noncanonical kinetic terms, i.e.\ 
$k$-essence \cite{kess1,kess2}, although this is often some function of the 
canonical kinetic term $X=-(1/2)g^{\mu\nu}\partial_\mu\phi\partial_\nu\phi$, 
i.e.\ $K(X)$. Here we explore what a pole structure may do in a dark energy 
context. Unlike inflation, dark energy is a late time phenomenon and has only 
had $N\approx1$ e-folds of influence. Similarly perturbations associated 
with dark energy tend to be negligible, at least on subhorizon scales 
(though $k$-essence theories can give more significant effects). Thus we 
don't expect an equivalent result to $n_s$--$r$ tracks, and are really 
just openly exploring what effects may arise. 

In Sec.~\ref{sec:map} we introduce the pole dark energy theory and 
examine the properties upon transformation to the canonical frame. 
We present illustrative numerical results for dynamical evolution of 
the field and the dark energy equation of state in 
Sec.~\ref{sec:numer}, identifying some interesting cases that have 
observational viability. Section~\ref{sec:swamp} speculates about the 
relation to swampland criteria, and Sec.~\ref{sec:concl} presents 
conclusions and further work.

\section{Kinetic Pole and Canonical Transformation} \label{sec:map} 

We begin with a scalar field Lagrangian with a pole in the kinetic 
term, and some potential $V(\sigma)$, 
\be 
{\mathcal L}=\frac{-1}{2}\frac{k}{\sig^p}(\partial\sig)^2-V(\sig)\ . 
\ee 
The pole can reside at $\sig=0$ without loss of generality, and has 
residue $k$ and order $p$. Poles can arise in theories due to 
nonminimal coupling to the gravitational sector, geometric properties of 
the K{\"a}hler manifold in supergravity, or as a signature of soft symmetry 
breaking (see, e.g., \cite{1507.02277,1602.07867} and references cited 
therein). Here we treat it phenomenologically. 

The kinetic term can be brought into canonical form 
\be 
{\mathcal L}=\frac{-1}{2}(\partial\phi)^2-V(\phi)\ , 
\ee 
by the transformation 
\bea 
\phi&=&\frac{2\sqrt{k}}{|2-p|}\,\sig^{(2-p)/2}\\ 
\sig&=&\left(\frac{|2-p|}{2\sqrt{k}}\right)^{2/(2-p)}\,\phi^{2/(2-p)}\ . 
\label{eq:sigphi} 
\eea  
We take the branch $\sig\ge0$ (note the field will not cross zero due to 
the pole). The pole dark energy Lagrangian now has the form 
\be
{\mathcal L}=\frac{-1}{2}(\partial\phi)^2-
V\left(\sig=\left(\frac{|2-p|}{2\sqrt{k}}\right)^{2/(2-p)}\,\phi^{2/(2-p)}\right)\ . 
\ee 

Note there is in general no exponential factor that stretches the potential 
and gives a flat plateau such as for $\alpha$-attractors. However, we will 
see some other interesting properties below. The case $p=2$ is a special 
case, and is the standard one used 
for such inflation (but see 
\cite{alpha3,1507.02277,1602.07867,kl}). In this instance we instead have 
\be 
\phi=\pm\sqrt{k}\ln\sig\ ,\qquad \sig=e^{\pm\phi/\sqrt{k}}\qquad (p=2)\ . 
\label{eq:sigphi2} 
\ee 
Here the exponential and stretching to give a flat plateau does appear. 
Dark energy with such 
$\alpha$-attractors was explored in, for example, \cite{alfde,akrami}, and 
we will not consider standard $\alpha$-attractor dark energy further, 
although we do discuss a new ``superattractor'' variant. 

We also note there are situations in which $V$ may not be regular at 
the pole (e.g.\ having a pole there itself), 
where conventionally a Maclaurin series is used such that 
$V\approx V_0-c\sigma$. 
Some theories such as supergravity generate the kinetic 
and potential terms from the same K{\"a}hler potential, and the coupling 
in coupled theories also enters in both terms, so both having poles 
is possible. See \cite{1602.07867} 
for further discussion. 

Let us now investigate how the transformation into the canonical kinetic 
term transforms the potential into a ``canonicalized'' potential, for 
various common initial potential forms. 
From the form of Eq.~(\ref{eq:sigphi}) we see that a power law potential 
is transformed into a power law potential, 
\be 
V\sim \sig^n \quad\longrightarrow\quad V\sim \phi^{2n/(2-p)}\ . 
\ee 
Note the characteristics depend on whether $p<2$ or $p>2$, in particular 
whether the initial and transformed power law indices are the same sign. 

For $p<2$ a monomial potential gets transformed 
into a monomial potential, and an inverse power law potential becomes 
an inverse power law potential. This is not particularly interesting, 
especially because it steepens the potential (e.g.\ for $p=1$ it takes 
$V\sim\sig^n$ to $V\sim\phi^{2n}$), making it less suitable 
for inflation or dark energy near cosmological constant like behavior 
(dark energy equation of state parameter $w\approx-1$). 
But for $p>2$, a monomial becomes an 
inverse power law and an inverse power law becomes a monomial, i.e.\ 
the signs flip. 

This is important since for 
canonical scalar fields, a monomial potential gives rise to thawing 
dark energy that starts in a cosmological constant like state at high 
redshift and evolves aways from it at later times, while an inverse power 
law gives freezing dark energy that can have a dynamical attractor 
behavior to a constant equation of state parameter $w$ at early times and 
then later evolves toward cosmological constant behavior \cite{caldlin}. 

Thus, pole dark energy can generate the properties of 
freezing, possibly attractor, fields from simple monomials (like 
$V\sim\sig^2$ or $\sig^4$), and thawing fields from 
an original inverse power potential. 
Recall that $\sig$ is defined on $[0,\infty]$, which is natural for an 
inverse power law potential and represents the positive field half of a 
monomial. Since $\phi$ also lives in $[0,\infty]$ the same holds for the 
transformed fields. 

This can lead to an interesting consequence, e.g.\ for potentials 
with a region of negative values. For example, with $p=2(1-n)$ this 
transforms an inverse power law into a 
linear potential $V\sim\phi$. However, the field never rolls past $\phi=0$ 
into the negative potential region since $\phi=0$ corresponds to 
$\sig=\infty$. Conversely, a linear potential in $\sig$ maps to an inverse 
power law dark energy for $p>2$, with $V\sim\phi^{-2/(p-2)}$. 

\begin{figure}[htbp!] 
\centering
\includegraphics[width=0.7\columnwidth]{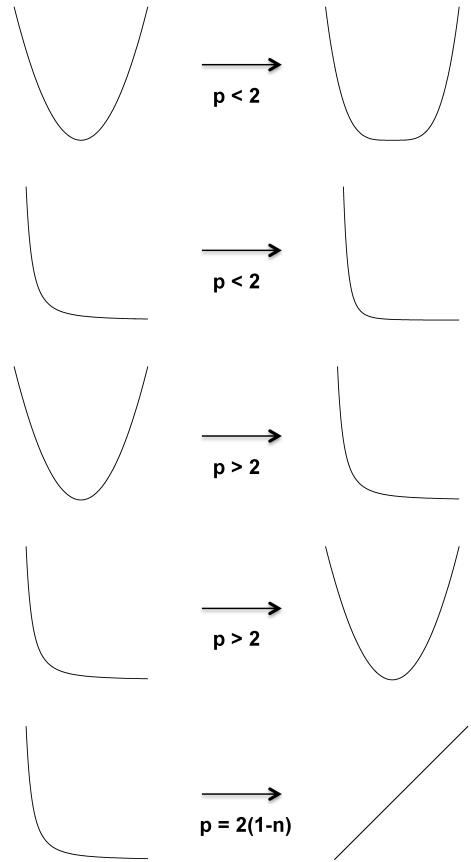} 
\caption{
The field transformation to give a canonical kinetic term from one with 
a pole of order $p$ also transforms the potential. Depending on the value 
of $p$ this can take monomials and inverse power laws into each other, 
including linear potentials. 
} 
\label{fig:vtransform}
\end{figure}

Figure~\ref{fig:vtransform} illustratively summarizes the mapping from 
power law and inverse power law potentials to the canonicalized potential 
form. Poles of order $p<2$ preserve the power law index sign (and make the 
potential steeper), while poles 
of order $p>2$ change the sign, turning a monomial into an inverse power 
law (and vice versa). The particular value $p=2(1-n)$ transforms an 
inverse power law into a linear potential.

One case of particular interest is $p=4$. This maps a power law index to 
its negative, i.e.\  $\sig^n\to\phi^{-n}$. For an exponential potential 
$V\sim e^{-\lambda\sig}$ becomes $V\sim e^{-\lambda\sqrt{k}/\phi}$, again 
mapping a classic dark energy freezer potential to a well known thawer. 

Another case of note is $p\gg1$. Not only will this convert a thawer to 
a freezer (since $p>2$) but it will take any monomial with index $m$ 
and bring it to an inverse power law potential with (negative) index 
$n\ll1$. The inverse power 
law characteristic gives the usual early time attractor dynamics, but the 
index $n\ll1$ provides $w_{\rm early}=(-2+nw_b)/(2+n)\approx -1+n(1+w_b)/2$, 
where $w_b$ is the background equation of state (e.g.\ $w_b=0$ during the 
matter dominated era. Thus $p\gg1$ generates close to 
cosmological constant like dynamics for any intrinsic monomial potential. 
Effectively, $p\gg1$ freezes the dynamics of $\sig$. This is not dissimilar 
to screening mechanisms in modified gravity where Vainshtein screening 
(or k-mouflage) works by decoupling the field through a large kinetic term 
\cite{screen1,screen2,screen3}.

\section{Numerical Dynamics} \label{sec:numer} 

To explore which theories, in terms of pole order and original form of 
potential, can usefully serve as dark energy, that is be observationally 
viable in the sense of having $w\approx-1$ at recent times (we will 
specifically use it in the sense that $w(z)<-0.9$), we consider 
some standard particle physics potentials and numerically solve the 
equations of motion of the transformed field.

\subsection{Quadratic $\to$ Inverse Power Law} \label{sec:n2ipl} 

Starting with a quadratic potential, i.e.\ $V=(1/2)m^2\sig^2$, which gives 
a thawing field, we choose $p>2$. If $p<2$ then the transformed potential 
will also be a monomial, and a steeper one, making it a less desirable 
dark energy model. For $p>2$ we will have an inverse power law 
canonicalized potential, which has the nice feature of an attractor 
behavior at early times during radiation and matter domination. 

\begin{figure}[htbp!] 
\centering
\includegraphics[width=\columnwidth]{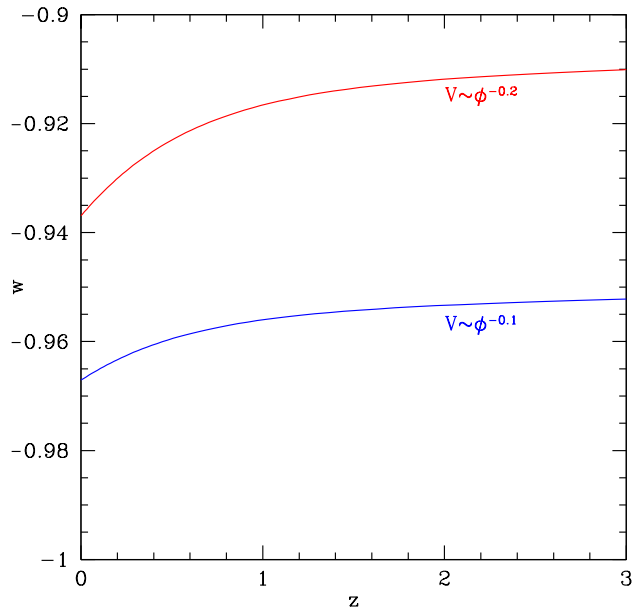} 
\caption{
The transformed inverse power law potential is observationally viable 
for power law index $\alpha=-2n/(2-p)\lesssim0.2$, with the dark energy 
equation of state parameter $w(z)<-0.9$. 
} 
\label{fig:iplw}
\end{figure}

However, inverse power law potentials $V\sim\phi^{-\alpha}$ have 
difficulty attaining $w\approx-1$ by the present, so they are not viable 
unless the power law index $\alpha\ll1$ (see, e.g., \cite{ratra}). 
We show the numerically obtained 
equation of state behavior $w(z)$ in Fig.~\ref{fig:iplw}. For 
$\alpha\lesssim0.2$ the dark energy evolution is roughly acceptable 
observationally. Note that it is a freezing model, with a high redshift 
constant $w$ as discussed in the previous section (with $w\to -0.909$, 
$-0.952$ for $\alpha=0.2$, 0.1) but approaching $w=-1$ as the universe 
evolves.

\begin{figure}[htbp!] 
\centering
\includegraphics[width=\columnwidth]{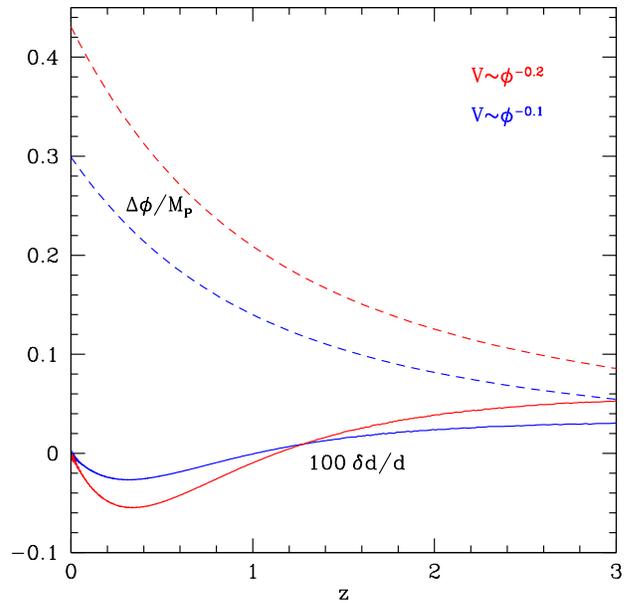} 
\caption{
The dark energy behavior from Fig.~\ref{fig:iplw} gives an expansion 
history similar to constant $w$ models, here $w=-0.963$ and $w=-0.929$ 
for $\alpha=0.1$, 0.2 (dark blue and light red curves) respectively. The 
fractional distance deviation from the corresponding constant $w$ model 
is shown 
(solid curves) in percent, i.e.\ the maximum deviation is 0.06\%. The 
field excursion $\Delta\phi$ is subPlanckian (dashed curves). 
} 
\label{fig:iplphid}
\end{figure}

Freezing fields were studied in detail in \cite{1701.01445}. In particular, 
for models not too far from $w=-1$ one can compare them to the constant $w$ 
dark energy defined by the calibration relation 
$w_{\rm const}=w(a_\star\approx0.85)$. Figure~\ref{fig:iplphid} 
demonstrates that indeed the distance observables in the inverse power law 
model and its partner constant $w$ agree at the 0.06\% level. Note that 
the field runs to the present over a subPlanckian excursion, $\Delta\phi<M_P$.

In pole dark energy one does not have to set the potential to have a 
shallow inverse power law index. The transformed potential $V\sim\phi^{-0.2}$ 
can be obtained from a quadratic potential $n=2$ with $p=22$, or a linear 
potential $n=1$ with $p=12$, or a monodromy potential $n=2/3$ with $p=8.67$.

\subsection{Dilation $\to$ Inverse Exponential} 

\begin{figure}[htbp!] 
\centering
\includegraphics[width=\columnwidth]{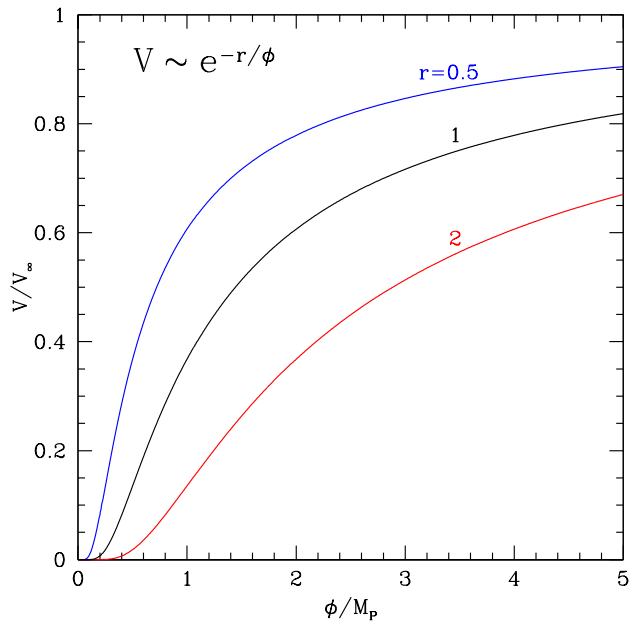} 
\caption{
The transformed inverse exponential potential $V\sim e^{-r/\phi}$ has 
a zero minimum and a plateau, like $\alpha$-attractors, but a weaker one. 
Nevertheless the field can stay frozen until nearly the present if it 
does not start too close to the minimum. 
} 
\label{fig:vinvexp}
\end{figure}

Another common particle physics potential is the exponential potential, 
$V\sim e^{-\beta\sig}$, such as from dilaton fields. This gives a freezing 
field, and indeed in 
the canonical case would trace the background energy density component 
evolution at early times. 
When we use this in a Lagrangian with a kinetic pole with $p=4$, 
we obtain $V\sim e^{-\beta\sqrt{k}/\phi}\equiv e^{-r/\phi}$. Such a 
potential gives thawing dark energy. The potential is shown in 
Fig.~\ref{fig:vinvexp} for three values of $r=\beta\sqrt{k}$.

Thawing fields were studied in detail in \cite{1501.01634}. In particular,
for models not too far from $w=-1$, the field excursion to the present follows 
\be 
\Delta\phi/M_P\approx 0.7\sqrt{1+w_0}\ . \label{eq:dphi} 
\ee 
We verify numerically that holds here. The potential is 
plateau-like at large $\phi$ (with $\phi=\infty$ corresponding to the 
pole at $\sig=0$), but not as flat as an $\alpha$-attractor. Note that 
near the minimum ($\phi=0$, corresponding to $\sig=\infty$, so the field 
cannot roll through zero) the potential is very flat, exponentially so, 
more than a monomial. 

\begin{figure}[htbp!] 
\centering
\includegraphics[width=\columnwidth]{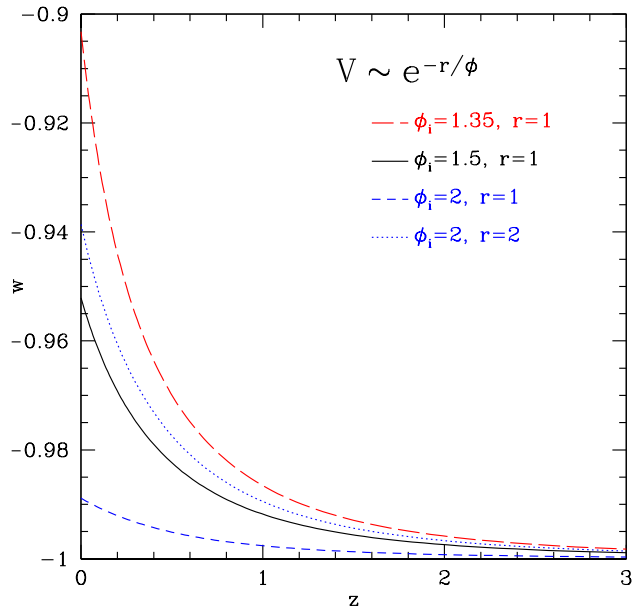} 
\caption{
The transformed inverse exponential potential $V\sim e^{-r/\phi}$ 
gives thawing dark energy, 
whose equation of state parameter today depends on the coefficient 
$r$ and the initial field value $\phi_i$. It is observationally viable 
if $\phi_i\gtrsim \sqrt{2r}$. 
} 
\label{fig:invexpwz}
\end{figure}

Figure~\ref{fig:invexpwz} shows the evolution $w(z)$. This can be 
observationally viable for parameters of order unity. Since the dark 
energy is a thawing model, it is sensitive to the initial field value 
$\phi_i$ at high redshift (the value $\phi_i$ is insensitive to the 
exact redshift 
used since the field is frozen until dark energy density becomes 
appreciable, e.g.\ $\Delta\phi/M_P<0.01$ until $\Omega_\phi\approx0.1$). 
The steepness of the potential $\lambda=-(1/V)dV/d\phi=-r/\phi^2$ helps 
determine how quickly the field thaws, and we find that the condition 
$\phi_i\gtrsim\sqrt{2r}$ (i.e.\ $|\lambda_i|\lesssim 1/2$) gives 
observational viability.

\subsection{Poles with $p=2$, and Superexponential} \label{sec:p2} 

We briefly return to the usual case of kinetic poles with $p=2$, as 
generally used for inflation. If we use a potential with a pole as well, 
such as an inverse power law $V\sim\sigma^{-n}$, then this is not 
covered by the usual Maclaurin expansion $V\approx V_0-c\sigma$. However, 
by Eq.~(\ref{eq:sigphi2}) this gives an exponential for the canonicalized 
potential, $V\sim e^{\mp\phi n/\sqrt{k}}$. Such a potential will not give 
a satisfactory late time acceleration, unless the field initial conditions 
are fine tuned to a thawing state (small kinetic energy) rather than the 
standard scaling solution. Note that like the $p<2$ case, the freezer 
inverse power law transforms to a freezer potential. 

A monomial potential, e.g.\ $V\sim\sig^2$, is the well known 
$\alpha$-attractor case (specifically the T model). 

Instead we will use a dilaton potential, $V\sim e^{-\beta\sigma}$, which 
gives rise to a superexponential behavior 
\be 
V\sim e^{-\beta e^{-\phi/\sqrt{k}}}\ . 
\ee 
Near the pole $\sigma=0$ ($\phi=\infty$) one can write this in the usual form 
\be 
V\sim 1-c\sigma \quad\longrightarrow\quad V\sim1-\beta\,e^{-\phi/\sqrt{k}}\ , 
\ee 
looking like an $\alpha$-attractor. However, away from the pole one can 
see the full transformed potential, and in fact the exponential of an 
exponential imbues it with a flatter plateau than a conventional 
$\alpha$-attractor, giving an enhanced ``basin of attraction'' for 
field initial conditions to yield $w\approx-1$ today. 

\begin{figure}[htbp!] 
\centering
\includegraphics[width=\columnwidth]{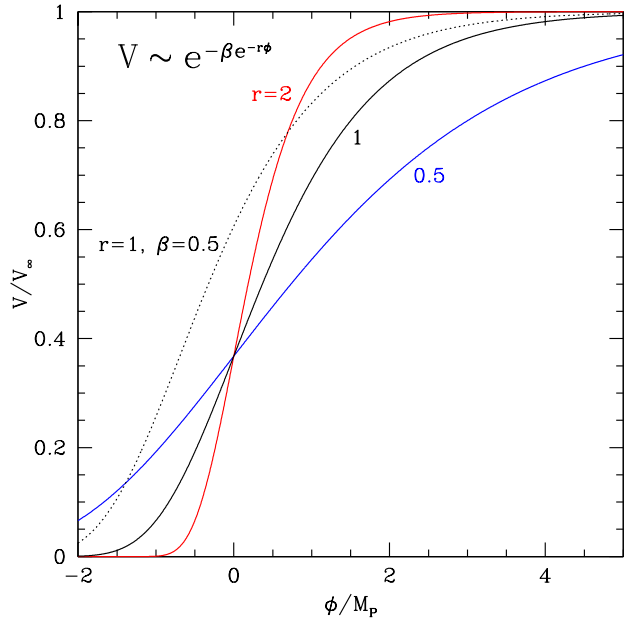} 
\caption{
The transformed superexponential potential 
$V\sim e^{-\beta\,e^{-r\phi}}$ has 
a zero minimum (at $\phi\ll -1/r$) and a plateau, like $\alpha$-attractors, 
but an enhanced one. Potentials are plotted vs $\phi$ for various values 
of $r$, with $\beta=1$ unless otherwise labeled. Increasing $r$ or 
decreasing $\beta$ enhances the plateau. 
} 
\label{fig:vsuper}
\end{figure}

We plot the transformed potential in Fig.~\ref{fig:vsuper}. Although 
$\sig$ is bounded by $[0,\infty]$, here $\phi$ can range over 
$[-\infty,\infty]$, with $\sig=\infty$ corresponding to $\phi=\infty$ 
(the potential plateau) and $\sig=0$ corresponding to $\phi=-\infty$ 
(the potential minimum at zero energy), while $\sig=1$ gives $\phi=0$.

\begin{figure}[htbp!] 
\centering
\includegraphics[width=\columnwidth]{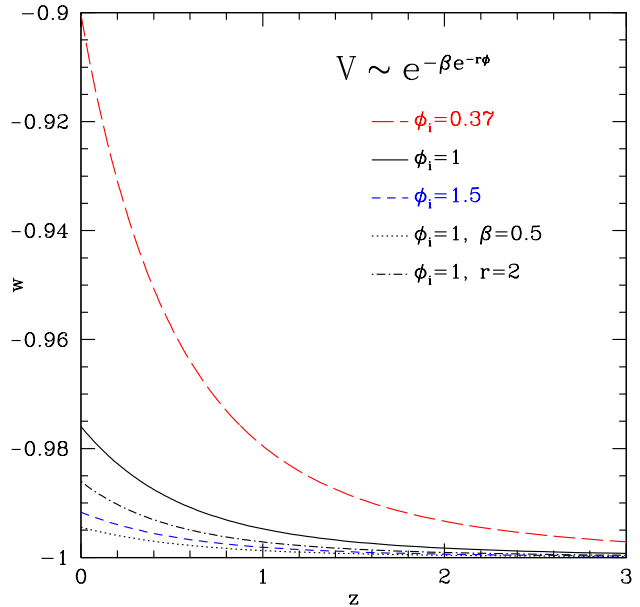} 
\caption{
The transformed superexponential potential 
gives thawing dark energy, 
whose equation of state parameter today depends on the coefficients 
$\beta$ and $r$ and the initial field value $\phi_i$. It is 
observationally viable for much lower $\phi_i$ than a conventional 
$\alpha$-attractor. The dark energy equation of state evolution $w(z)$ 
is plotted for various $\phi_i$, with $r=1$, $\beta=1$ unless otherwise 
labeled. The enhanced potential plateau from increasing $r$ or decreasing 
$\beta$ leads to a less thawed $w(z)$, one remaining closer to $-1$. 
} 
\label{fig:superwz}
\end{figure}

Note that increasing $r=1/\sqrt{k}$ or decreasing $\beta$ strengthens the 
plateau and the attraction to $w\approx-1$. Figure~\ref{fig:superwz} 
shows the dark energy equation of state evolution. Even initial field 
values as low as $\phi_i/M_P=0.37$ (for the $r=1$, $\beta=1$ case), can give 
observationally viable dark energy behavior -- a much smaller value than 
for most $\alpha$-attractors (for example the Starobinsky model requires 
$\phi_i/M_P\gtrsim1.5$, and an ordinary nonattractor like a quadratic 
potential requires $\phi_i/M_P\gtrsim2.5$, as seen in Fig.~2 of 
\cite{alfde}). 
One clearly sees that increasing $r$ or decreasing $\beta$ keeps $w(z)$ 
closer to $-1$. Since this is a thawing field, it again follows the 
field excursion amplitude given by Eq.~(\ref{eq:dphi}), and so is 
subPlanckian. (For notational convenience we will sometimes set $M_P=1$.)

\section{Can Poles be Stilts over the Swamp?} \label{sec:swamp} 

Scalar field potentials frequently used for dark energy models can have 
difficulty being consistent with string theory, in terms of such 
conditions as swampland criteria \cite{1610.01533,1806.08362}. These 
conjecture that there is a distance limit 
$\Delta\phi/M_P\lesssim{\mathcal O}(1)$ 
and a steepness criterion $|\nabla V|/V\gtrsim {\mathcal O}(1)$. 

These conditions impose severe constraints on the types of potentials 
usually favored for inflation or dark energy. However, we can ask whether 
the use of poles in the kinetic term can relax the requirements on the 
potential. This has been mentioned for inflation in terms of noncanonical 
kinetic terms and the distance criterion in \cite{1807.05445} and for 
both criteria within multifield inflation in \cite{1807.04390}. In fact, 
in the latter the field space metric gives an effective noncanonical 
kinetic term. For inflation, changing the kinetic term can give rise to 
a tension between satisfying swampland criteria and 
nongaussianity constraints \cite{1806.09718}. 

Dark energy does not have observational nongaussianity limits so it seems 
worthwhile to explore whether the noncanonical kinetic term used here can 
ease swampland constraints. In effect, does the pole enable the field to 
go over the swampland? The answer in a formal sense is no, but in a 
practical sense is possibly. 

For the first criterion, we have already discussed that the field excursions 
for the models considered, in the observationally viable part of parameter 
space, follow Eq.~(\ref{eq:dphi}) for the thawing fields, giving 
$\Delta\phi/M_P<1$. The freezing model of Sec.~\ref{sec:n2ipl} also 
obeyed that limit, as shown in Fig.~\ref{fig:iplphid}. Of course in 
the future the field will travel further, but the condition holds in the 
observable region, the past. 

Regarding the steepness criterion, this can also be satisfied under certain 
conditions. For a noncanonical kinetic term $K(\sigma)(\partial\sigma)^2/2$, 
this takes the form 
\cite{1806.08362,1807.04390} 
\be 
\frac{|\nabla V|}{V}=\frac{1}{\sqrt{K}}\,\frac{|dV/d\sig|}{V}=\sqrt{\frac{\sig^p}{k}}\,\frac{|dV/d\sig|}{V}\ . 
\ee 
Indeed, this works out the same as first canonicalizing the field and 
then simply using $|dV/d\phi|/V$. Nevertheless, our canonicalized potentials 
are unusual enough that it is worth calculating. 

For our first model, a monomial canonicalizing to an 
inverse power law potential, we have 
\be 
V\sim \sig^n \quad;\quad \frac{|\nabla V|}{V}= 
\frac{|n|}{\sqrt{k}}\,\sig^{(p-2)/2}=\frac{2|n|}{|2-p|}\frac{1}{\phi} \ . 
\ee 
Since $p>2$ we need $\sig\gtrsim1$ or $\phi\lesssim1$. At early times 
$\phi$ can indeed start small and we see we get viable dark energy for 
subPlanckian excursions so it is possible that the steepness is of 
order unity, possibly satisfying the steepness criterion. 

For the second model, transforming from an exponential to an inverse 
exponential potential, we have 
\be 
V\sim e^{-\beta\sig} \quad;\quad \frac{|\nabla V|}{V}=\beta 
\sqrt{\frac{\sig^p}{k}}\to\frac{\beta\sqrt{k}}{\phi^2}\ , 
\ee 
for our value $p=4$. 
We can satisfy the criterion for large $r=\beta\sqrt{k}$, but 
this is observationally 
unviable since it drives $w$ far from $-1$. Again we want 
$\sigma\gtrsim1$ or $\phi\lesssim1$. As we see in 
Fig.~\ref{fig:invexpwz}, this makes it more difficult to achieve 
viable $w(z)$, and this model is not as satisfactory in avoiding 
the swampland. 

For the third model, an exponential canonicalizing to a superexponential, 
the same condition on $\sig$ holds, 
\be 
V\sim e^{-\beta\sig} \quad;\quad \frac{|\nabla V|}{V}=\beta
\sqrt{\frac{\sig^p}{k}}\to\frac{\beta}{\sqrt{k}}\,e^{-\phi/\sqrt{k}}\ , 
\ee 
although the condition is different in terms of $\phi$ (here with 
$p=2$). 
Again we would like $\beta$ large, but this is observationally unviable, 
or $k$ small (i.e.\ $r$ large), which is ok, but we must make sure that 
$\phi/\sqrt{k}$ is not large. 
Fig.~\ref{fig:superwz} demonstrates that these conditions give viable 
$w(z)$ (as long as $\phi$ isn't too small, well off the plateau toward 
the minimum), and so we can potentially achieve $|\nabla V|/V\sim 1$. 

Thus at least some of these pole dark energy models seem to have an 
acceptable range in which they could satisfy both swampland criteria 
and observational viability 
in the form of $w(z)<-0.9$ for all redshifts to the present.

\section{Conclusions and Further Thoughts} \label{sec:concl} 

Noncanonical kinetic terms can arise from a wide variety of physics, from 
Dirac-Born-Infeld  to higher dimension to coupled models. They add a 
degree of freedom to quintessence, allowing a dark energy sound speed 
lower than the speed of light and dynamics in a distinct region of phase 
space, give nongaussianity in inflation, and enable a type of screening in 
modified gravity theories. Poles in the kinetic term can arise from several 
physics mechanisms, and be tied to underlying geometric considerations 
in string theory. Here we explored the impact of kinetic poles on dark 
energy and cosmic acceleration in the recent universe. 

The transformation from the noncanonical kinetic term into the 
canonical term, which in $\alpha$-attractor models leads to an 
extended plateau suitable for inflation (or dark energy), can also 
turn freezing dark energy into thawing dark energy, and vice versa. 
We study different order poles and their effect on standard 
potentials, demonstrating that they can deliver viable dark energy 
models. In many cases the evolution can approach $w\approx-1$ 
more easily than in the standard canonical case. 

Three example models we study are an inverse power law potential 
with very small index, generated from a standard quadratic potential 
(so a thawer transformed to a freezer), an inverse exponential 
generated from a dilaton field (a freezer transformed to a thawer), 
and a superexponential coming from a dilaton field. This last model 
shows an enhanced plateau, flatter than an $\alpha$-attractor, and 
superattraction in terms of a significantly expanded basin of attraction 
toward $w\approx-1$ relative to monomials and even $\alpha$-attractors. 
We also discuss how a high order pole $p\gg1$ enables cosmological 
constant like behavior independent of potential, acting in a manner 
somewhat analogous to Vainshtein screening in modified gravity. 

We numerically test all these models for evolution that is observationally 
viable, in the specific sense that $w(z)<-0.9$. Moreover, all the models 
exhibit a subPlanckian field excursion (up to the present). 

Finally, we consider a more speculative question of whether pole dark 
energy can avoid swampland criteria difficulties. We have already seen 
that it satisfies the distance criterion. We demonstrate that indeed the 
pole models can potentially obey the steepness criterion, yet be 
observationally viable dark energy with $w\approx-1$. Whether this 
opens fruitful avenues for dark energy within string theory is left for 
future work. 

The poles in the noncanonical 
theories lead to interesting new dark energy models such as the 
superexponential one that exhibits superattraction in the sense of a 
significantly expanded set of field initial conditions that give rise to 
observationally viable evolution, and new methods of attaining 
interesting old dark energy models like a very shallow inverse power 
law attractor model. Thus pole dark energy appears worthy of 
future investigation.

\section*{Acknowledgments}

I thank KASI for hospitality during part of this work.  
This work is supported in part by 
the U.S.\ Department of Energy, Office of Science, Office of High Energy 
Physics, under Award DE-SC-0007867 and contract no.\ DE-AC02-05CH11231, 
and by the Energetic Cosmos Laboratory.

\end{document}